# Predicting Large Hadron Collider Observations using Kazuo Kondo's Mass Quantum Cascade

Grenville J. Croll
grenville@spreadsheetrisks.com

**The late Kazuo Kondo left a hitherto unknown *a priori* particle theory which provides predictions of massive particles which may be detected by the Large Hadron Collider (LHC). This article briefly introduces Kondo's work and documents the derivation and masses of his expected hyper-mesons, hyper-hadrons, heavy leptons and massive neutrinos. Several particles in these classes may have already been detected.**

**1.0 Introduction**

Almost every month for almost three decades until his death aged 90 in December 2001, Professor Emeritus Kazuo Kondo, Dept. Applied Mathematics, Tokyo University, Japan, published a monograph developing his ideas on his mathematically based natural philosophy. There are 358 monographs in total [Kondo, 1973-2001], each monograph comprising 16-48pp (referenced herein as K1-K358). These monographs continue earlier work started in the 1950's prior to his formal retirement from Tokyo University in 1973 [Kondo 1955, 1958, 1962, 1968]. Kondo's personal output, estimated at four to five million words, is one of the largest bodies of scientific work completed by a single individual. More remarkable is the content [Croll, 2006] and the fact that it has lain undiscovered and unappreciated for so long.

Kondo's philosophy presupposes that reality consists of information elements, in myriad permutations and combinations and that they are indistinguishable, finite in number and countable. That is to say, in Kondo's view, there is no underlying physical reality, merely number. Reality is perceived through the counting of number. There is a mapping between Kondo's integral parameter space and the continuous three dimensional space we perceive through a mathematical construct known as the Kawaguchi space. The Kawaguchi space was first referred to as such by Synge [Synge, 1935] in the early part of the twentieth century.

Kawaguchi spaces are a class of higher order space. The properties of such spaces have been studied widely. Miron [Miron, 1998] has recently produced a number of standard texts on higher order geometry, but the bulk of the work on Kawaguchi spaces in particular was completed in Japan in the first half of the twentieth century [Kawaguchi, A, 1931, 1932, 1933] [Kawaguchi, M., 1962 & 1968]. Two critically important results, the KH and KHK theorems, due to Kawaguchi, Hombo [Kawaguchi, 1941, P94, Footnote 2] and later Kondo [K291, P14] prove that for Kawaguchi spaces of higher order, the dimensionality of such spaces need not exceed three (the other dimensions cancelling out due to cross-differentiation). Kondo's interpretation of these theorems, which he discusses repetitively, is that they are the *a priori* mathematical reason why reality has the dimensionality that we observe.

The existence of pre-spaces as a model of reality has been previously postulated by Hiley [Hiley, 2000] and others. The pre-space model of reality is appealing because of the simplicity of representation of the complex structures within reality such as molecules, crystals and biological forms etc. The Kawaguchi transformation facilitates the mapping of just a few numeric parameters into complex spherical and other topological continuous forms.

The use of the Kawaguchi space is further appealing in that there is a natural mathematical axiom: particles are spaces and spaces are particle, reflecting the wave particle duality of quantum mechanics. Kondo's use of Kawaguchi's mathematics is regularly reconciled to conventional Quantum Mechanics [Kondo, 1958a] [Kondo, 1997] [K336]. Note that Kondo's work is not another interpretation of Quantum Mechanics as the familiar quantum mechanical formalisms arise naturally out of the higher order geometry. For example, Kondo's Quantum-Mechanical Approximation Lemma [ Kondo, 1997, p28] states:

> *"The non-relativistic Schrödinger wave mechanics is an approximation of the Zermelo conditions valid within a microscopic domain of size O(dx)"*.

The boundary between the macroscopic and microscopic is clearly defined by consideration of the dimensionality of the Kawaguchi spaces representing objects. Zero and positive dimensionality corresponds to macroscopically observable items. Negative dimensionality corresponds to the microscopic, unobservable or quantum mechanical world.

In one of the last recasts of his wide ranging ideas, in his ninth decade [K316], Kondo addresses the need to include incompleteness within his mathematically consistent model of reality. Kondo's finite integral parametric pre-space, though mathematically consistent, is of necessity incomplete due to Gödel's theorem. Completeness is achieved by equating the pre-space with an infinite identity, in a manner highly reminiscent of the famous identity $e^{i\pi} = -1$. On the right hand side of the equation is rationality (ie arithmetic), finity and incompleteness. On the left are irrationality, infinity and completeness.

Latterly, he states in his Galois Field Recognition Lemma[K316, p3]:

> *"The epistemological recognition, or consciousness, of human beings, must start with the extraction of Galois fields"*.

Thus we are given, through Kondo's insight and Kawaguchi's mathematics, a set of physical theories which incorporate at their heart the essence of consciousness.

Kondo's work is a varied mix of abstract mathematics and extended discussion across a very wide variety of topics including nuclear physics, gravity, relativity, chemistry, biology, optics, information theory, plant and animal taxonomy, physiology, neurophysiology, psychology, musicology and linguistics. Running through all the work is the persistent use of the Kawaguchi space as the mathematical model which appears to underpin the natural world.

Kondo does not ignore the contemporary view, including the Standard Model. Very many of his discussions [eg K336] address the commonalities and differences between his own ideas and those of contemporaries such as Einstein, Dirac, Bohr etc. The use of higher order geometry appears to give a richer and more widely applicable theory.

Kondo applies his theories to particles detected over the years by various scientific apparatus. The fit is not yet perfect. Importantly, a careful reading of his last works [Kondo, 1997][K303-5][K347] reveals a reasonably clear set of predictions of higher energy particles likely to be detected by equipment such as the LHC.

Having introduced Kondo and his work, this paper provides firstly an extremely terse introduction to Kondo's low mass/energy particle model and secondly an outline of his explicit expectations with regard to higher mass/energy observations. References are provided to the original monographs which are definitive in so far as they can be given the rapid development of his ideas across his many monographs. The majority of the rest of this paper is due to Kazuo Kondo, whether referenced, acknowledged or not.

**2.0 The Kawaguchi Space**

The following is extracted from "Three Phases of Epistemological Penetration to Nature" [Kondo, 1997, page 2]:

...in the following exposition we shall first show that the fundamental constitution of epistemological recognition has to be put in terms of expoint co-ordinates. They are a systematical means of comparison of information built up in the domain of real numbers.

Expoints are expressed by sets of terms of co-ordinates:

$$x^i, i=1, 2, \ldots N$$

and parameters

$$\mu^\lambda, \lambda = 1, 2, \ldots L$$

by the expoint coordinates

$$x^i_{\lambda(r)} = \frac{\partial^r x^i}{\partial x^i_{\lambda 1} \ldots \partial x^i_{\lambda r}} \quad r = 0, 1, \ldots M$$

or, in the simpler case of L =1

$$x^{(r)i} = \frac{d^r x^i}{dt^r} \quad r = 0, 1, \ldots M$$

The manifold of expoints is called a space. It is shown that the significant phases are restricted to the space configurations under the Zermelo-Géhéniau conditions in terms of $x^i_{\lambda(r)}$, which space we indicate by $K^{(M)}_{LN}$ [i.e. For **M** Derivatives, **N** Dimensions and **L** Parameters]. In the single parametric case, restricted by the Zermelo conditions, the space is indicated by $K^{(M)}_{N}$.

**3.0 The Construction of Elementary Particles**

The following is extracted from K293 of November 1995:

**Definition I.** A particle is carried on the Higher Order Space $K^{(M)}_{LN}$ One can say that the space is itself a particle and the particle is itself a space.

The constitution and the behaviour of the particle are, therefore, restricted by the Zermelo-Géhéniau conditions.

**Definition II.** A specific kind of particle carried on the line element of $K^{(M)}_N$ is called a **lepton**. The leptons are restricted by the Zermelo conditions.

**Definition III.** Mesons are of the next kind of particle given for L = 2. A more complicated class given for L = 3 are **baryons**. Both of baryons and mesons are collectively called **hadrons.**

If L is odd, L-parametric particles are fermions else they are bosons.

**3.1 Leptons**

The lepton is constrained by the Zermelo condition equation:

$$\Delta_1 F \equiv ( x^{(1)i} \partial_{(1)i} + \kappa ) F = F$$

where

$$\kappa = \sum_{r=2}^{M} r\, x^{(r)i} \partial_{(r)i}$$

represents the so-called **mass** of the particle and

$$\Theta_r = x^{(r)i} \partial_{(r)i}$$

works as a mass quantum.

The lepton of order M = 1 is the neutrino, having a mass of $\approx 0$.

The mass quanta obtained from the differential equations of the single parametric space corresponding to the lepton are denoted $\theta$ and $\Theta$. $2\theta$ obtaining from order M = 2 and $3\Theta$ from order M = 3. As M = 3 is the KH limit, mass contributions from higher orders of M are equivalent and held to be insignificant for low lying (i.e. lower energy) leptons. This suggests that:

$$m_e = 2\theta, \quad m_u = 2\theta + 3\Theta$$

and compared with the experimental masses

$$m_e = 0.511 \text{ Mev}, \quad m_u = 105 \text{ Mev}$$

gives

$$\theta = 0.2555 \text{ Mev}$$

and

$$\Theta = 35 \text{ Mev}$$

from which follows, approximately, the familiar ratio:

$$\alpha = \theta / \Theta \approx 1 / 137$$

i.e. the fine structure constant is the ratio of the mass quanta associated with orders M=1 and M=2. For consistency with what follows, we tabulate lepton structure:

## Table 0: Lepton Structure

| | | | | | | |
|---|---|---|---|---|---|---|
| a) | Neutrino | [ \| ] | | > | 0* | |
| b) | Electron/positron | [ 2\| ] | $2\theta$ | = | 0.511 | Mev |
| c) | Muon/antimuon | [ 2\|3 ] | $2\theta + 3\Theta$ | = | 105 | Mev |

*  Mass arises from a lower quantum – see later.

### 3.2 Hadrons

The hadron is constrained by the Zermelo-Géhéniau conditions. The multiparametric features of $\mathbf{K}^{(M)}{}_{LN}$ can be described in terms of L phases $u$ each associated with one of the L Parameters:

$$\Delta_u{}^{\beta(H)} = \sum_{r=H}^{M} \binom{r}{H} x^i_{u\gamma(r-H)} \frac{\bar{\partial}}{\partial_{\beta(H)\gamma(r-H)}}$$

$$= \Delta_H(u) + \Delta^*{}_H$$

Where $\Delta_H(u)$ has the same structure as the Zermelo operator $\Delta_H$ on the arc line with parameter $u$ and $\Delta^*{}_H$ is responsible for the interactions between the constructions of the other parametric lines. Thus for hadrons, mass contributions arise through inter-parametric (inter-quark) interferences, represented by the mass quanta $\Theta_{u\gamma(r-1)}$ where $u\gamma(r-1)$ refers to more than one quark. Thus some quark mass arises through inter quark (i.e. interparametric) interference, neatly explaining quark confinement.

For baryons, by letting the three parameters (quarks) be denoted by $u, v, w,$ the following kinds of mass quanta are enumerated:

| | | | | |
|---|---|---|---|---|
| $\Theta_u,$ | $\Theta_v,$ | $\Theta_w$ | = | $\Theta$ |
| $\Theta_{uu},$ | $\Theta_{vv},$ | $\Theta_{ww}$ | = | $2\Theta$ |
| $\Theta_{uv},$ | $\Theta_{vw},$ | $\Theta_{wu}$ | = | $2\Theta$ |
| $\Theta_{uuu},$ | $\Theta_{vvv},$ | $\Theta_{www}$ | = | $3\Theta$ |
| | $\Theta_{uuv},$ | $\Theta_{uvv}$ | = | $3\Theta$ |
| | $\Theta_{vvw},$ | $\Theta_{vww}$ | = | $3\Theta$ |
| | $\Theta_{wwu},$ | $\Theta_{wuu}$ | = | $3\Theta$ |
| | | $\Theta_{uvw}$ | = | $3\Theta$ |

Which gives rise to Kondo's baryon structure summarised in Table 1:

## Table 1 Baryonic Structure

| representative type | Θu | Θuu | Θuuu | Θuuv | Θuv | Θuvw |
|---|---|---|---|---|---|---|
| *Multipliers* | 1 | 2 | 3 | 3 | 2 | |
| | 1 | 2 | 3 | 3 | 2 | 3 |
| | 1 | 2 | 3 | 3 | 2 | |
| | reduced to Θ | reduced to Θ | | θ Ignored | | θ Ignored |

The quanta located to the left of ‖ comprise Θ only. The quanta located to the right of ‖ (with mixed indices) are composed of Θ's and θ's, the mass of the latter being presently ignored due to their small contribution (θ << Θ).

The mass quanta $\Theta_{uv}$, $\Theta_{vw}$, $\Theta_{wu}$ fall on ε(2) meaning an element of the principal ideal (2) generated by 2 in the ring of integers so that the multiplier can be any even number or zero. The Mass quanta $\Theta_{uuv}$, $\Theta_{uvv}$, $\Theta_{vvw}$, $\Theta_{wwv}$, $\Theta_{wwu}$, $\Theta_{wuu}$ fall on an element ε(3) of the principal ideal (3) generated by 3, so that the multiplier can be any integral multiple of 3 or zero. The quantum $\Theta_{uvw}$ also falls on ε(3) Θ, in particular on the triplet Θ + Θ + Θ.

Since the macroscopic observer does not recognise the distinction between the positive and negative directions of the parameters *u, d, s,* corresponding to the three rows (quarks), an invariance needs to be maintained such that the mass construction must be left right symmetric.

There is no *a priori* reason why quarks should not manifest with terms of order M = 4 (or more) ie with a space of $\mathbf{K}^{(4)}_{3,3}$ which gives rise to the possibility of a further term of magnitude 4 appearing in one or more of the rows above. These are *strange* quarks, with part of their structure lying outside the KH limit of M = 3. Table 2 enumerates the basic flavours of baryon. Note that terms of *θ* are ignored. The notation of Table 2 is based on the structure of Table 1.

**Table 2: Basic Baryonic composition, showing possible quark composition and neglecting lower mass quanta**

a) N: $\begin{bmatrix} 1\ 2\ 3\ .\ .\ 2\ 1 \\ 1\ 2\ 3\ .\ .\ 2\ 1 \\ 1\ 2\ 3\ .\ .\ 2\ 1 \end{bmatrix}$     27 Θ = 945 Mev,

b) Σ: $\begin{bmatrix} 1\ 2\ 3\ .\ .\ 2\ 1 \\ 1\ 2\ 3\ .\ .\ 2\ 1 \\ 1\ 2\ 3\ 4\ 3\ 2\ 1 \end{bmatrix}$     34 Θ = 1190 Mev,

c) Ξ: $\begin{bmatrix} 1\ 2\ 3\ .\ .\ 2\ 1 \\ 1\ 2\ 3\ 4\ .\ 2\ 1 \\ 1\ 2\ 3\ 4\ 3\ 2\ 1 \end{bmatrix}$     38 Θ = 1330 Mev,

a') Δ: $\begin{bmatrix} 1\ 2\ 3\ .\ 3\ 2\ 1 \\ 1\ 2\ 3\ .\ 3\ 2\ 1 \\ 1\ 2\ 3\ .\ 3\ 2\ 1 \end{bmatrix}$     36 Θ = 1260 Mev,

b')    Y*:    $\begin{bmatrix} 1\,2\,3\,.\,3\,2\,1 \\ 1\,2\,3\,.\,3\,2\,1 \\ 1\,2\,3\,4\,3\,2\,1 \end{bmatrix}$    40 Θ   =   1400 Mev,

c')    Ξ*:    $\begin{bmatrix} 1\,2\,3\,.\,3\,2\,1 \\ 1\,2\,3\,4\,3\,2\,1 \\ 1\,2\,3\,4\,3\,2\,1 \end{bmatrix}$    44 Θ   =   1540 Mev,

d')    Ω:    $\begin{bmatrix} 1\,2\,3\,4\,3\,2\,1 \\ 1\,2\,3\,4\,3\,2\,1 \\ 1\,2\,3\,4\,3\,2\,1 \end{bmatrix}$    48 Θ   =   1680 Mev,

e)    Λ:    $\begin{bmatrix} 1\,2\,.\,3\,.\,2\,1 \\ 1\,2\,.\,3\,.\,2\,1 \\ 1\,2\,3\,2\,3\,2\,1 \end{bmatrix}$    32 Θ   =   1120 Mev,

Counting the number of strange quarks (those that include the numeral 4), the items Δ, Y*, Ξ*, Ω are said to have strangeness 0, -1, -2, -3. The items N, Σ, Ξ are said to have strangeness 0, -1, -2. The Λ particle is said to have a weak strange quark falling on a mass of 2 Θ with strangeness of -1. The negative sign of the strangeness numbers is related to the sign of the electric charge accorded to each quark. The above structures are used to also derive Spin, Charge, Hypercharge and Isospin.

Note that the following celebrated mass formulae [Coleman & Glashow (1961)] [Gell-Mann, 1962] apply exactly given Kondo's basic particle composition of Table 2:

$$m_\Omega - m_\Delta = 3(m_{\Xi^*} - m_{Y^*})$$

$$m_\Xi - m_\Sigma = m_{\Xi^*} - m_{Y^*}$$

$$m_\Sigma - m_\Lambda + 2/3\ (m_N - m_\Xi) = 2/3\ (m_\Delta - m_{\Xi^*})$$

$$(m_N + m_\Xi)/2 = (3m_\Lambda + m_\Sigma)/4$$

As does:

$$4(m_\Lambda - m_N) = 10\ (m_\Sigma - m_\Lambda) = 5\ (m_\Xi - m_\Sigma)$$

Mass differences within the hadron submultiplets are accounted for by tracing influences of the lower mass quantum θ (0.255 Mev). The difference between the neutron and the proton could be 5 θ = 1.275 Mev ($m_n$ - $m_p$=1.293 Mev). The mass difference between the hyperons could be 19 θ = 4.845 Mev ($m_{\Sigma^-}$ - $m_{\Sigma 0}$ = 4.88 +/- 0.008Mev). The mass difference within the Ξ submultiplet could be 24 θ = 6.12 Mev ($m_{\Xi^-}$ - $m_{\Xi 0}$ = 6.4 +/- 0.6 Mev).

**3.3 Mesons**

Mesons are two parametric (L = 2) particles containing two quarks. Their summary structures, particle associations and masses (the mass to the left of | corresponds to θ) are given in Table 3 below.

## Table 3: Outline Meson Structure

a)    $\eta$    $\begin{bmatrix} 1 & |2\,3\,2\,1 \\ 1 & |2\,3\,2\,1 \end{bmatrix}$    $16\,\Theta\ =\ 560\ \text{Mev},$

b)    $\rho$ or $\omega$    $\begin{bmatrix} 1 & |2\,3\ \ 3\,2\,1 \\ 1 & |2\,3\ \ 3\,2\,1 \end{bmatrix}$    $22\,\Theta\ =\ 770\ \text{Mev},$

c)    $K$    $\begin{bmatrix} 1 & |2\,3\ \ 2\,0 \\ 1 & |2\,3\ \ 2\,0 \end{bmatrix}$    $14\,\Theta\ =\ 490\ \text{Mev},$

d)    $K^*$    $\begin{bmatrix} 1 & |2\,3\ \ \ \ \ 3\,2\,1 \\ 1 & |2\,3\ 4\,3\,2\,1 \end{bmatrix}$    $26\,\Theta\ =\ 910\ \text{Mev},$

e)    $\eta'$    $\begin{bmatrix} 1 & |2\,3\ \ 2\,3\,2\,1 \\ 1 & |2\,3\ 4\,3\,2\,1 \end{bmatrix}$    $28\,\Theta\ =\ 980\ \text{Mev},$

f)    $\varphi$    $\begin{bmatrix} 1 & |2\,3\ 4\,3\,2\,1 \\ 1 & |2\,3\ 4\,3\,2\,1 \end{bmatrix}$    $30\,\Theta\ =\ 1050\ \text{Mev},$

g)    $\pi$    $\begin{bmatrix} 1 & |2 \\ 1 & |2 \end{bmatrix}$    $4\,\Theta\ =\ 140\ \text{Mev},$

Note that the Okubo mass rule

$$m^2_K = ( 3\, m_\eta^2 + m_\pi^2 ) / 4$$

applies exactly as do the simpler relations:

$$m_{K^*} - m_{\rho,\omega} = m_\varphi - m_{K^*}$$

$$6(m_K - m_\pi) = 30\,(m_\eta - m_K) = 5(m_\eta - m_{\eta'})$$

Note also that contributions from the lower mass quantum $\theta$ are ignored.

## 4.0 The Ultimate Microscopic Mass assembly

Following a detailed homo/cohomological argument given in K291, Kondo states in K293 that "..*the ultimate microscopic mass assembly must be at any rate of three dimensional shape and of the simplest kind. It cannot but be a three-dimensional simplex having tetrahedral form*". Based on its homological structure, the tetrahedron can carry the following mass quanta:

    4 Vertices carry 4 unnamed lower quanta having mass $\theta_0$
    6 Edges carry 3 Nucleons having mass $\theta_1$
    4 Faces carry 4 Kaons having mass $\theta_2$
    1 Interior carries 1 further unnamed lower quanta having mass $\theta_3$

$\theta_2$    $\begin{bmatrix} 2\,3\,2 \\ 2\,3\,2 \end{bmatrix}$    $= 14\,\Theta$

$$\theta_1 \quad \begin{bmatrix} 1\,2\,3\,2\,1 \\ 1\,2\,3\,2\,1 \\ 1\,2\,3\,2\,1 \end{bmatrix} = 27\,\Theta$$

At a first approximation, the tetrahedron has six edges to provide 3 pairs of supports for the 3 Nucleon equivalents to carry $3 * 27\,\Theta$. It has four triangular faces providing supports for 4 Kaon equivalents to carry $4 * 14\,\Theta$.

In all the tetrahedron carries

$$3 * 27\,\Theta + 4 * 14\,\Theta = 81\Theta + 56\,\Theta = 137\,\Theta = \P = 4795 \text{ Mev}$$

on this basic mass assembly unit. See figure 1.

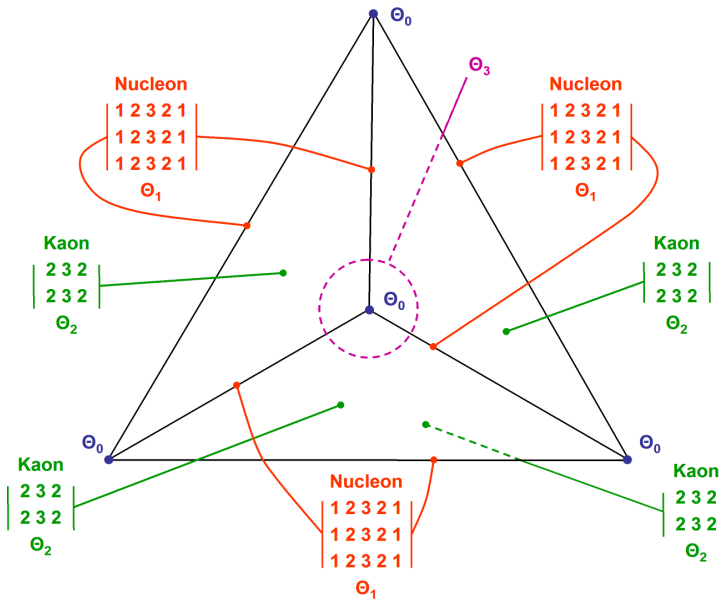

*Figure 1 – Kondo's Tetrahedral Mass Assembly*

At a better approximation for the tetrahedron, Kondo suggests that the vertexes are equipped to carry 4 units of a certain lower kind of quanta $\theta_0$ and the interior is equipped to carry another mass unit, say $\theta_3$, giving a total mass carried of

$$\P = 4\,\theta_0 + 56\,\theta_2 + 81\,\theta_1 + \theta_3$$

Kondo postulates that

$$\theta_1 = \theta_2 = \Theta \quad \text{and} \quad \theta_0 = \theta_3 = \theta$$

In which case

$$\P = 5\,\theta + 137\,\Theta$$

substituting $\alpha = \theta / \Theta$, $\alpha = \Theta / \P$

$$1 = 5\,\alpha^2 + 137\,\alpha$$

Solving the quadratic for $\alpha$ yields

$1/\alpha = 137.036$ (4793)

giving a better approximation to the fine structure constant.

Kondo discusses the composition of the tetrahedron, and the variation expected in the fine structure constant at higher energies when tetrahedra are more closely packed. It is suggested that edge and face quanta can be shared leading to reductions in the fine structure constant to $\alpha = 1/128$, $\alpha = 1/129$ and $\alpha = 1/127$, as has been observed in the high energy region with regard to the behaviour of particles $W^{\pm b}$ etc.

**5.0 Cascade of Mass Quanta**

The above introduces a cascade of mass quanta based on $\Theta = \theta / \alpha$ which Kondo extends in both directions:

$\theta''$ = $\theta' \alpha$     etc

$\theta'$ = $\theta \alpha$ = 0.015 Mev

$\theta$ = $\Theta \alpha$ = 0.255 Mev,

$\Theta$ = $\theta / \alpha$ = 35 Mev ,

¶ = $\Theta / \alpha$, = 4.795 Gev

¶' = ¶ $/ \alpha$ = 656   Gev

¶'' = ¶' $/ \alpha$   etc

Kondo expects that the structure of the low lying hadronic world (based on $\Theta$) can be nearly isomorphically mapped into a higher level where $\Theta$ is replaced by ¶, ¶' etc.

Evidence for these hyper-particles is not absent. So called *fire balls* and *andromedas* have been traced from cosmic rays by emulsion chamber experiments in the Andes. Quanta around 2~3, 10~20 and 100~300 times the mass of the nucleon have been reported. They are called respectively

   *HQ (hyper quanta),*       *SHQ (super-hyper-quanta)*

   *UHQ (ultra – hyper – quanta )*

in the report [Chacaltaya, 1971].

The discovery of bosons at mass levels around 10 Gev mostly indicated by Υ suggests a massive quark of around ¶. It is suggested that the so-called top quark, lying in the region 175 Gev, corresponds to a particle of around 36¶.

There is no suggestion from Kondo that the tetrahedral mass assembly may well be recursive such that the obscure quantum $\theta_3$ associated with the centre region of the tetrahedral mass assembly of Figure 1 is also tetrahedral but constructed using lower mass quanta $\theta$. Likewise, the structure of the ¶' very heavy mass quantum may well be tetrahedral, recursively containing a ¶ heavy mass quantum at its centre.

## 6.0 Hyper Mesons & Hyper Hadrons

Although Kondo states that the higher energy regime is isomorphic to the Hadrons, he only enumerates the hyper mesons, naming them so. His hyper meson table, slightly abbreviated and orthogonally transposed[Kondo, 1997], is given below.

### Table 4 Abbreviated Hyper-Meson Table

| Class | Symbol | Particle Mass (Gev) | Quark Mass (Gev) | Dominant Quantum | Fire Ball | Quark |
|---|---|---|---|---|---|---|
| Meson | $\varphi$ | 1.05 | 0.525 | $\Theta$ | | s |
| Intermediate | D | 2.0 | 1.0 | $\Theta$ | HQ | s,c |
| Intermediate | J/$\psi$ | 3.097 | 1.5 | $\Theta$ | HQ | s,c |
| Hyper | $\pi/\alpha$ | 19.18 | 9.59 | ¶ | SHQ | u/$\alpha$, d/$\alpha$ |
| Hyper | $\eta/\alpha$ | 76.72 | 38.36 | ¶ | UHQ | u/$\alpha$, d/$\alpha$ |
| Hyper | $\rho/\alpha$ | 105.49 | 57.745 | ¶ | UHQ | u/$\alpha$, d/$\alpha$ |
| Hyper | $\varphi/\alpha$ | 143.85 | 71.92 | ¶ | UHQ | s/$\alpha$ |
| Hyper | D/$\alpha$ | 274.0 | 137.0 | ¶ | UHQ | s/$\alpha$, b/$\alpha$ |
| Hyper | J/$\psi$/$\alpha$ | 421.96 | 210.98 | ¶ | UHQ | b/$\alpha$, t |

The hyper mesons of Table 4 may be observed in the LHC, with by extension, a similar set of particles based on the ¶' quantum. The hyper hadrons are isomorphic to the hadrons of Table 1 and are based on ¶, so that for example, a hyper nucleon of mass 27 ¶ = 656.9 Gev is expected.

## 7.0 Heavy Gauge Bosons

It is repeatedly pointed out that the W and Z gauge bosons quantise at almost exactly 17¶ = 81.5 Gev and 19¶ = 91.1 Gev. The approximate construction of a family of heavy gauge bosons is postulated as per Table 5:

### Table 5 Simpler Heavy Gauge Bosons

a) $\begin{bmatrix} 1\ 2\ 3\ |\ 2\ 0 \\ 1\ 2\ 3\ |\ 0\ 0 \end{bmatrix}$    14 ¶ = 67.12 Gev,

b) W    $\begin{bmatrix} 1\ 2\ 3\ |\ 2\ 3 \\ 1\ 2\ 3\ |\ 0\ 0 \end{bmatrix}$    17 ¶ = 81.48 Gev,

c) Z    $\begin{bmatrix} 1\ 2\ 3\ |\ 2\ 0 \\ 1\ 2\ 3\ |\ 2\ 3 \end{bmatrix}$    19 ¶ = 91.09 Gev,

d) $\begin{bmatrix} 1\ 2\ 3\ |\ 2\ 3 \\ 1\ 2\ 3\ |\ 0\ 3 \end{bmatrix}$      20 ¶    =    95.88 Gev,

The quanta to the right of | can vary according to which elements of the principle ideals (2) and (3) come to into play. Thus the W particles would appear to be hyper Kaons and the Z particle one half of a hyper zeta'. The primality of the mass configurations for the more readily observed W & Z bosons is noted, as is their propensity to appear in pairs.

Note that the so called top quark mass of approximately 36 ¶ equals the mass of the hypothesized hyper-delta particle, and that decomposition into W & Z particles would seem feasible.

The author speculates that a family of ultra heavy gauge bosons could be expected with masses between 14 ¶' = 9.179 Tev and 20 ¶' = 13.13 Tev, just under the LHC expected performance of 14 Tev, though unlikely to be observed by it.

## 8.0    Higgs Mass

The following is taken directly from Kondo's Accademia Pontaniana article [Kondo, 1997, page 61], though it appears elsewhere in the Post Raag reports.

In a phenomenological model proposed by Weinberg and Salam [Weinberg, 1967] [Salam, 1968][Weinberg; Salam; Glashow; 1980], the masses of the W and Z are expressed by

$$m_W = gv/2, \qquad m_Z = gv/2\cos\theta_w$$

where

$$g = e \sin\theta_{w,}, \qquad \cos\theta_w = 0.8944$$

referring to the electric charge unit $e$, and

$$v = 264 \text{ Gev}$$

However, no *a priori* reason is included as to why $\theta_w$ and $v$ should fall on these specific numerical values, even if $g$ is supported by the assumption of an electro weak unification and there is a certain reason that $v$ is related to the Fermi coupling constant $G_F$ by

$$\sqrt{2}\ G_F = v^{-2}$$

We offer instead an a priori reason that $m_W$ and $m_Z$ are theoretically fixed as above so that

$$\cos\theta_w \quad = \quad 17/19 \quad = \quad 0.8994737$$

$G$ and $v$ being also fixed accordingly. The Fermi constant can be calculated therefrom. The *a priori* theoretical and the phenomenological experimental values are compared in the following table

The above is summarised in Kondo's Electroweak Unification Lemma: [K293, p108] *"A unified theory of electroweak phenomena is given in terms of the geometry of higher order spaces through the Zermelo-Geheniau conditions, the characteristic constants e, g, g, v, $\theta_w$ all being theoretically introduced and evaluated"*.

| | $\theta_w$ | $v$ | $G_F$ |
|---|---|---|---|
| | Table 6 | $\theta_w \sim v$ Comparison Table | |
| *H.O.S. Theoretical* | 26.525 | 260.42 Gev | 1.2333 E-5 Gev-2 |
| *Empirically Based* | 26.565~30 | 264.00 Gev | 1.6639(2) E-5 Gev-2 |

Obviously $gv = 37¶$ and the Fermi constant are transforms of the inverse square of the hyper-mass quantum ¶ which will turn out to be the Higgs mass.

## 9.0 Heavy Leptons and Massive Neutrinos

The hyper-quantum ¶ cannot exist in an isolated condition for long and will go through a rapid decomposition. The most likely decomposition is into two major subunits:

One consisting of $81 \Theta = 3 * 27 \Theta$ =   2.835  Gev

One consisting of $56 \Theta = 4 * 14 \Theta$ =   1.96   Gev

However of the original $56 \Theta$, 3 times $17 \Theta = 51 \Theta$ may appear corresponding to 3 W particles that are well defined in the ¶ regime. Causing $5\Theta$ to associate with $81 \Theta$ giving another possible subunit of $86 \Theta = 3.010$ Gev (J/ψ) and (by implication) $51 \Theta = 1.785$ Gev. The $\Theta$ then decompose into $\theta$ leaving residues of

|   |   |   |
|---|---|---|
| | $86 \theta$ =  | 21.93 Mev |
| | $81 \theta$ =  | 20.65 Mev |
| And | | |
| | $56 \theta$ =  | 14.28 Mev |
| | $51 \theta$ =  | 13.00 Mev |

The particles as outlined above are expected to be observed during heavy quantum (¶) decay. The lighter of which ( < 24 Mev) is or are massive neutrinos, conventionally denoted ντ. The heavier particles are heavy leptons, one of which, the τ has been reported to have been observed at 1.776 (+0.30-0.27) Gev.

The above decay process is $¶ = \Delta\tau + \Delta v_\tau$ where the left hand side goes into nothing or an obscure trace at the disintegration.

## 10.0  Summary

We have extracted from Kondo's four to five million published words some elements which apply to low and high energy particle physics. We have outlined the origin of mass which is obtained from Akitsugu Kawaguchi's geometry of higher order spaces. We have described the structure of leptons and hadrons in terms of a cascade of mass quanta derived from the higher orders of this geometry. The cascade of mass quanta are hypothesized to be organised into a recursive tetrahedral geometry which provides clues about the disintegration routes of heavy mass quanta. Sets of hyper mesons and hyper hadrons are proposed based upon the simple isomorphic substitution of heavy mass quanta for the lighter mass quanta thought to be associated with presently observed phenomena. Heavy leptons, heavy neutrinos and heavy bosons are also proposed. The origin of the fine structure constant is derived from the higher order structure of the electron and muon. A reason behind the variation of the fine structure constant at higher energies is proposed which provides for further possible variation of the observed masses of heavy particles. An epistemological derivation of electro-weak unification is outlined and related to the origin of the Higgs Mass. A suggestion is given that the top quark is the hyper delta particle.

## 11.0 Further Work

It is hoped that copies of Kondo's post RAAG monographs on particle physics, particularly those referenced herein, can be transcribed and uploaded to ArXiv as soon as possible.

## Acknowledgements

The author acknowledges the generosity of Professor Tomoaki Kawaguchi in providing the author with a near complete set of the Post RAAG monographs (weighing over 20kg). This paper would not have been possible without Professor Kawaguchi's generosity. I thank Professor Laxmi Chandra Jain for copies of his 40 year correspondence with Professor Kondo. I thank my colleagues at ANPA for their encouragement and Dr Keith Bowden, TPRU Birkbeck College (ANPA President), for his review of this paper. This paper is dedicated to the memory of Kazuo Kondo.

## References

Chacaltaya (1971) Chacaltaya Emulsion Chamber Experiment and related papers, Supplements of the Progress of Theoretical Physics, No 47

Coleman, S., Glashow, S. L., (1961) Electromagnetic Properties of Baryons and the Unitary Symmetric Scheme, Physical Review Letters, 6, 423-425.

Croll G.J. (2006), The Natural Philosophy of Kazuo Kondo. Proc ANPA, Cambridge, arXiv:0712.0641v1 [math.HO]

Gell-Mann, M., (1962) Strange Particle Physics, Strong Interactions, Proceedings of the International Conference for Higher Energy Physics, CERN

Glashow, S.L., (1980) Towards a Unified Theory: Threads in Tapestry, Review of Modern Physics, Vol 52, No.3 (1980) 539-543

Hiley, B.J., (2000), Non-Commutative Geometry, the Bohm Interpretation and the Mind-Matter Relationship, Proc. ANPA 22 14-29

Kawaguchi, A., (1931, 1932, 1933), Theory of Connections in the Generalised Finsler Manifold, I, II, III, Proceedings of the Imperial Academy, Tokyo, Vols 7, 8, 9

Kawaguchu, A., (1941), Die Differentialgeometrie höheren Ordnung III: Erweiterte Paramatertransformationen un P-Tensoren, Journal of the Faculty of Science, Hokkaido Imperial University, 40, 77-150;

Kawaguchi, M., (1962), An Introduction to the Theory of Higher Order Spaces I: The Theory of Kawaguchi Spaces, RAAG Memoirs, Vol III 718-734

Kawaguchi, M., (1968) An Introduction to the Theory of Higher Order Spaces II: Higher Order Spaces in Multiple Parameters, RAAG Memoirs, Vol IV, 577-591

Kondo, K., Ed. (1955, 1958, 1962, 1968) RAAG (Research Association for Applied Geometry) Memoirs, Volumes 1,2,3,4, Gakujutsu Bunken Fukuy-Kai, Tokyo

Kondo, K., Ed (1958a), On the Fundamental Principle of Observation and Quantum-Mechnical Formalisms, RAAG Memoirs, Vol. II E-II, pp252-284

Kondo, K., (1973-2001), Post RAAG Reports Numbers 1-358: Research Notes and Memoranda of Applied Geometry for Prevenient Natural Philosophy, CPNP, 1570 Yotsukaido, Yotsukaido City, Chiba-ken, 284 Japan.

Kondo, K., (1997), Three Phases of Epistemological Penetration to Nature, Quaderni Dell'Accademia Pontaniana, No 20, Napoli

K291, Kondo, K., (1995) Post RAAG report Number 291, A Unified System of Epistemological Penetration to Nature, Part I: Introduction and Fundamentals

K293, Kondo, K., (1995) Post RAAG report Number 293, A Unified System of Epistemological Penetration to Nature, Part III: Construction of Elementary Particles

K303-5, Kondo, K., (1996) Post RAAG report Numbers 303-5, Mass Spectrum of Elementary Particles and Quarks, Part I: Introduction, Part II: Mass Spectrum at the GKH Limit, Part III : Mystery of Hyper-Mass-Quanta

K316, Kondo, K., (1997) Post RAAG report Number 316, Perception of Space, Matter and Bio-Psychological Information, Part I: Galois Field, Multidimensional Space and Differential Geometry

K336, Kondo, K., (1999) Post RAAG report Number 336, On Some Epistemological Questions Concerning Macroscopic and Microscopic Recognitions, Part I: Why the Fundamental Recognitions Fall on the Relativistic and Quantum Mechanical Formalism

K347, Kondo, K., (2000) Post RAAG report Number 347, Some Crucial Questions on Elementary Particles

Miron, R., (1998) *The Geometry of Higher-Order Finsler Spaces.* Hadronic Press

Salam, A., (1968) in Elementary Particle Theory, Proceedings of the 8th Nobel Symposium, edited by N. Svartholm, Almqvist & Wiksal, Stockholm, 361-..

Salam, A., (1980) Gauge Unification of Fundamental Forces, Review of Modern Physics, Vol 52, No 3, 511-538;

Synge, J.L., (1935) Some Intrinsic and Derived Vectors in a Kawaguchi Space. American Journal of Mathematics, 57, pp679-691

Weinberg, S., (1967) Physical Review Letters, 19, 1264.

Weinberg, S., (1979) Review of Modern Physics, Vol 52, No 3 (1980), 515-523